\documentclass[a4paper,11pt]{article}

\usepackage[square,numbers,sort&compress]{natbib}
\usepackage{graphicx}
\usepackage{amsmath}
\usepackage{amssymb}

\newcommand{\ket}[1]{\left |#1\right\rangle}
\newcommand{\eh}{\epsilon_{H}}

\newcommand{\Imax}{K_\text{H}}
\newcommand{\Iacc}{K_\text{acc}}
\newcommand{\fln}{f}

\title{\Large\bf Measuring the internal state of a single atom without energy exchange}

\author{J\"urgen Volz$^1$, Roger Gehr$^1$, Guilhem
  Dubois$^{1,2}$,\\J\'er\^ome Est\`eve$^1$
 \& Jakob Reichel$^1$\\
\normalsize $^1$Laboratoire Kastler-Brossel, ENS, CNRS, UPMC,\\
\normalsize 24 rue Lhomond,  75005 Paris, France.\\
\scriptsize  $^2$Present address: Astron FIAMM, 83210 La Farl{\`e}de, France
}

\date{May 17, 2011}

\begin{document}

\maketitle

%\begin{abstract}
{ \bf Real quantum measurements almost always cause a much stron\-ger
  back action than required by the laws of quantum mechanics. Quantum
  non-demolition (QND) measurements have been
  devised \cite{Braginsky92,Grangier98,Nogues99,Maioli05,Hume07,Lupascu07}
  such that the additional back action is kept entirely within
  observables other than the one being measured. However, this back
  action to other observables often also imposes constraints. In
  particular, free-space optical detection methods for single atoms
  and ions such as the shelving technique \cite{Leibfried03}, though
  being among the most sensitive and well-developed detection methods
  in quantum physics, inevitably require spontaneous scattering, even
  in the dispersive regime \cite{Hope05}. This causes irreversible
  energy exchange and heating, a limitation for atom-based quantum
  information processing where it obviates straightforward reuse of
  the qubit. No such energy exchange is required by quantum
  mechanics \cite{Kwiat95}. Here we experimentally demonstrate optical
  detection of an atomic qubit with significantly less than one
  spontaneous scattering event.
%and we experimentally quantify the amount
  We measure transmission and reflection of an optical
  cavity \cite{Boozer06,Puppe07,Khudaverdyan09,Bochmann10} containing
  the atom. In addition to the qubit detection itself, we
    quantitatively measure how much spontaneous scattering
  has occurred.  This allows us to relate the
  information gained to the amount of spontaneous emission, and we
  obtain a detection error below 10\% while scattering less than 0.2
  photons on average. Furthermore, we perform a quantum Zeno type
  experiment to quantify the measurement back action and find that
  every incident photon leads to an almost complete state collapse.
  Together, these results constitute a full experimental
  characterisation of a quantum measurement in the ``energy
  exchange-free'' regime below a single spontaneous emission
  event. Besides its fundamental interest, this means significant
  simplification for proposed neutral-atom quantum computation
  schemes \cite{Ladd10} and may enable sensitive detection of molecules
  and atoms lacking closed transitions.}

In the first step of a measurement, the system to be measured becomes
entangled with another quantum object (``meter''), such as a photon
field. For the case of a two-level system (qubit),
$(\alpha\ket{0}+\beta\ket{1})\otimes\ket{\Psi_{\text{in}}}$ evolves
into $\alpha\ket{0}\otimes
\ket{\Psi_0}+\beta\ket{1}\otimes\ket{\Psi_1}$.  The readout of the
qubit then amounts to distinguishing the meter states $\ket{\Psi_0}$
and $\ket{\Psi_1}$, which can only be achieved up to some error
because they are generally nonorthogonal. The minimum possible
detection error $\epsilon=(\epsilon_0+\epsilon_1)/2$ is given by the
Helstrom bound \cite{Helstrom76}
\begin{equation}
 \epsilon_H = \frac{1}{2}\left(1-\sqrt{1-\left|\langle \Psi_0|\Psi_1\rangle\right|^2}\right),\label{eq:eqn1}
\end{equation}
where $\epsilon_0$ and $\epsilon_1 $ are the probabilities to measure
the qubit in $\ket{1}$ although it was in $\ket{0}$ and vice versa and
we assume no prior knowledge on the qubit state. In the following, we
consider the generic case where a qubit is probed by
an incident coherent light pulse containing $n$ photons on average. To
good approximation, the two final states $\ket{\Psi_0}$ and
$\ket{\Psi_1}$ then also consist of coherent states.  As an example,
consider an ideal fluorescence measurement in which the dark state
$\ket{0}$ does not interact with the light, while the bright state
$\ket{1}$ scatters all photons.  In this case,
$\ket{\Psi_0}=\ket{0}_S\ket{n}_T$ and
$\ket{\Psi_1}=\ket{n}_S\ket{0}_T$, where $\ket{n}$ is a coherent pulse
containing $n$ photons on average, and $S$ and $T$ refer to the
scattered and transmitted light modes. Then,
$\left|\langle\Psi_0|\Psi_1\rangle\right|^2=\exp(-2 n)$, and in the
limit of large $n$ one obtains $\eh \approx \exp(-2n)/4$. More
generally, in all schemes using coherent pulses (so that
$\ket{\Psi_0}$ and $\ket{\Psi_1}$ are tensor products of coherent
states each containing a photon number proportional to $n$),
$\left|\langle \Psi_0|\Psi_1\rangle\right|^2=\exp(-\zeta n)$ with
some real $\zeta$.  This exponential decrease of the minimum
detection error with $n$ naturally leads to a heuristic definition of
the {``knowledge''} on the atomic state as {$K\equiv-\ln2\epsilon$}. The
maximum knowledge {$\Imax$} one can obtain from a measurement is then
%$\Imax=\fln \equiv -\ln 2\epsilon_H$.
$\Imax= -\ln 2\epsilon_H$.  We use the notation $\fln(x)= -\ln
\left(1- \sqrt{1- \exp(-x)}\right)$, which for large $x$ simplifies to
$\fln(x) \approx x$. Thus, {for coherent pulse schemes,}
${\Imax=}\fln(\zeta n)$ is the knowledge that the environment has
obtained during the measurement, and constitutes an upper bound to the
knowledge $\Iacc$ that the experimenter can actually access. In the
case of the ideal fluorescence measurement, the maximum knowledge is
$\fln(2n) \approx 2n$.

The measurement leads to a back action on the atom. The final state of
the qubit after the measurement is obtained by tracing over the
meter \cite{Haroche06}: the coherence of the qubit is reduced,
$\rho_{0,1}\to \langle\Psi_0|\Psi_1\rangle\rho_{0,1}$, where $\rho$ is
the qubit density matrix. This intrinsic back action need not
affect other degrees of freedom (DOFs) of the system: DOFs which do
not get entangled with the meter can be factored out and remain
unaffected by the measurement.  Most real measurements, however, cause
a much larger back action. In particular, fluorescence measurements are
inevitably accompanied by spontaneous emission, which leads to
heating and may pump the atom to an internal state outside the qubit
basis. In the example of an ideal fluorescence detection, each
incident photon is spontaneously scattered when the atom is in the
bright state. Therefore, in terms of scattered photons $m$, the {maximum
knowledge can be expressed as
\begin{equation}
\Imax=\fln(2m)\approx 2m\,.
\label{eq:spbound}
\end{equation}}
This bound in fact applies to all free-space measurement methods in
which classical light sources are used in a single-pass
configuration \cite{Hope05}: in all such methods, information gain is
necessarily accompanied by energy exchange between the atom and the
light. In particular, this includes dispersive measurements with far
off-resonant light. Moreover, even state-of-the-art experiments
typically fall short of this limit by several orders of magnitude due
to limited collection efficiency and require the scattering of a large
number of photons to infer the qubit state. \cite{Gerber09}

\begin{figure}[htb]
  \includegraphics[width=\columnwidth]{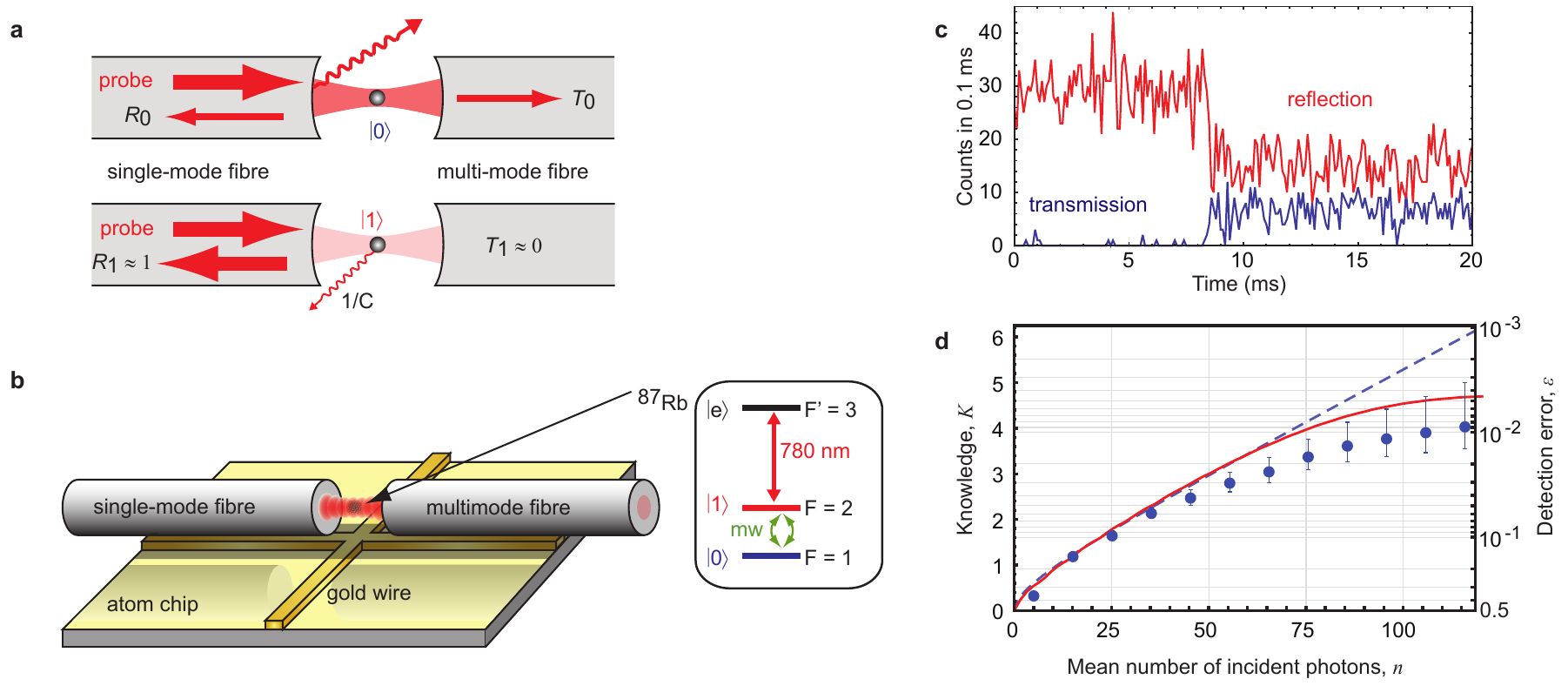}
  \caption{\label{fig:setup} \small\textbf{Cavity-assisted detection
      of an atomic qubit.} \textbf{a}, For an atom in the dark state
    $\ket{0}$, probe light is either transmitted, reflected or lost by
    mirror imperfections. For the bright state $\ket{1}$, most
    incident photons are reflected. In both cases, only a small
    fraction is scattered by the atom. \textbf{b}, Our cavity is
    formed by the coated end facets of two optical fibres. The qubit
    states ($F=1$, $m_F=0$ and $F=2$, $m_F=0$) can be coupled by a
    resonant microwave. Cavity and the atomic transition
    $\ket{1}\rightarrow\ket{e}$ are resonant to the $\pi$-polarised
    probe laser at 780 nm.  \textbf{c}, Typical detection trace
    showing cavity transmission (blue) and reflection (red) for an
    atom initially in $\ket{1}$ performing a quantum jump to $\ket{0}$
    due to spontaneous emission. \textbf{d}, Detection error and
    corresponding knowledge on the atomic state with one s.d. error
    bars versus incident photon number $n$. The dashed line is the
    theoretical prediction taking into account our cavity
    imperfections (see text). We exclude the possibility of quantum
    jumps during the measurement which explains the deviation for
    large $n$. The solid line is the full simulation of our detection
    process including quantum jumps.}
\end{figure}

We overcome this limit by coupling the atomic qubit to a cavity in the
strong-coupling regime $C=g^2/2\kappa\gamma\gg1$, where $g$ describes
the coherent atom-cavity coupling and $\kappa$ ($\gamma$) is the
cavity (atomic) decay rate.
The cavity is resonant to an optical transition of the $\ket{1}$
state, and probed by a resonant light pulse
(Fig.~\ref{fig:setup}\,a). An atom in the non-resonant state $\ket{0}$
has negligible effect on the cavity and all photons from the incident
mode are transmitted, $\ket{\Psi_0}=\ket{0}_R\ket{n}_T$. By contrast,
an atom in $\ket{1}$ detunes the cavity by more than its linewidth, so
that almost all photons are reflected,
$\ket{\Psi_1}\approx\ket{n}_R\ket{0}_T$. The states $\ket{\Psi_0}$ and
$\ket{\Psi_1}$ thus have the same overlap as in the ideal fluorescence
measurement, and $\Imax=\fln(2n)$ as before. Now,
however, the atom sees a significant light intensity only when it is
in the non-resonant state. Quantitatively, the $\ket{1}$ state only
scatters a fraction $1/C$ of the incident photons
 \cite{Lugiato84,Hechenblaikner98}. Therefore, the {maximum knowledge} per
\emph{scattered} photon is $C$ times larger than the free-space limit:
\begin{equation}
\Imax=\fln(2Cm) \approx 2Cm\,.
\label{eq:spbound2}
\end{equation}
Furthermore, in contrast to fluorescence measurements,
$\ket{\Psi_0}$ and $\ket{\Psi_1}$ are modes that are easily experimentally accessible.
The atomic state can therefore be inferred with negligible spontaneous emission in a realistic experimental setup.

Our implementation of this cavity readout scheme is shown in
Fig.~\ref{fig:setup}\,b. The key element is a fiber-based high-finesse
cavity \cite{Colombe07,Hunger10b} {($g=2\pi$ $185\pm8$ MHz,
  $\kappa=2\pi$ $53\pm0.5$ MHz, $\gamma=2\pi$ 3 MHz, $C=108\pm8$)}
mounted on an atom chip.  We prepare a Bose-Einstein condensate of
$^{87}$Rb atoms, from which we load a single atom into an intracavity
dipole trap \cite{Gehr10}. Levels and transitions are shown in
Fig.~\ref{fig:setup}\,b.  As shown earlier \cite{Gehr10}, cavity
transmission and reflection allow us to efficiently read out the
atomic qubit state (see Fig.~\ref{fig:setup}\,c). Compared to the
ideal situation described above, our system suffers from mirror losses
and from the presence of the second TEM00 cavity mode with orthogonal
polarisation detuned by 540\,MHz, which has a
non-negligible coupling to the atom (see methods). Losses reduce the
empty-cavity transmission to $T_0=0.13$ with associated reflection
$R_0=0.42$. The presence of the second mode together with the effect
of the hyperfine atomic structure degrades the extinction ratio of the
transmission when a resonant atom is present.  Instead of the ratio
$T_1/T_0=1/(4 C^2)$ expected for a single-mode cavity coupled to a
two-level atom, we measure $T_1/T_0=2\%$ (see
Fig.~\ref{fig:setup}\,c), where $T_1=0.0024$ ($R_1\approx 1$) is
the cavity transmission (reflection) with an atom in the
resonant state $\ket{1}$. The two field states $\ket{\Psi_0}$ and
$\ket{\Psi_1}$ now have additional components for the loss channels
and for the outgoing modes coupled to the second cavity mode. This
increases $\langle\Psi_0\ket{\Psi_1}$, leading to a reduced
$\Imax=\fln(0.62n)$ (see methods). However, the intensity inside the
cavity is also reduced by the mirror loss and therefore the expected
knowledge in terms of scattered photons is still much higher than
$m$ (see methods).

Knowledge on the atomic state carried by photons lost at the mirrors
is not accessible to the experimenter, reducing the available
knowledge to $\fln(0.23n)$. Furthermore, photon counting in
the reflected and transmitted modes is not an optimal strategy to
distinguish the two states $\ket{\Psi_0}$ and $\ket{\Psi_1}$. The
associated detection error {$\epsilon_d$} is given by the overlap
of the two probability distributions of reflected and transmitted
counts detected when the atom is in $\ket{0}$ or $\ket{1}$.  As
predicted by the {Chernoff bound} \cite{Chernoff52}, it decreases
exponentially for large $n$ with a rate $\xi$, where $\xi$ can be
calculated from the reflection and transmission coefficients of the
cavity and the efficiency of the photon detectors (see
methods). Therefore, in the large-$n$ limit, the {knowledge that
can be experimentally accessed $\Iacc = -\ln 2\epsilon_d$}
follows the function $\fln(x)$ introduced above, with $x=\xi n$.
We numerically checked that $\Iacc \approx \fln(\xi n)$ is also a
valid approximation for small $n$. Taking into account finite photon
detection efficiencies ($47\%$ in transmission, $31\%$ in reflection),
we expect our detection method to yield $\Iacc =
\fln(4.6\times10^{-2}n)$ (which would increase to $\fln(0.11n)$ using
perfect detectors). To verify this prediction, we prepare the atom in
either of the qubit states $\ket{0}$ and $\ket{1}$ and measure the
corresponding detection errors \cite{Gehr10}. For $n<40$, the
measurement is in good agreement with the prediction (figure
\ref{fig:setup}\,d).  For larger $n$, nonresonant excitation leads to
a small probability of depumping the qubit from its initial state
during the measurement \cite{Gehr10}, thereby reducing the accessible
knowledge in the experiment.

\begin{figure}[htb]
\includegraphics[width=\columnwidth]{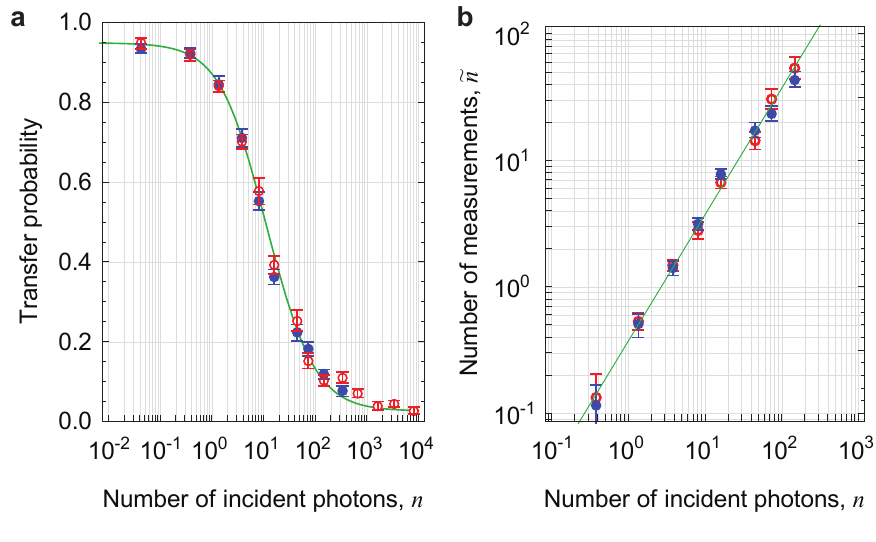}
\caption{\label{fig:Zeno} \small\textbf{Back action measurement using
    a quantum Zeno effect.} A microwave $\pi$-pulse (duration $8.8$
  $\mu$s) is applied to an atom in $\ket{1}$ ($\bullet$) or $\ket{0}$
  ($\circ$) in the presence of measurement light. \textbf{a}, Data
  points (with one s.d. error bars) show the transfer efficiency
  versus number of photons incident on the cavity during the
  pulse. \textbf{b}, The average number of projective measurements
  $\tilde{n}$ we deduce for each data point from our model; the solid
  line is a linear fit $\tilde{n}=a_0 n$ yielding $a_0=0.37\pm0.02$.
  The solid line in \textbf{a} shows the prediction of our
  theoretical model supposing this linear relation and value of
  $a_0$.}
\end{figure}

Although we only detect part of the incident photons and therefore of
$\Imax$, we can still quantify $\Imax$ by its back action. We perform
the following quantum Zeno \cite{Itano06} experiment: An atom is
prepared in state $\ket{0}$ or $\ket{1}$ (see methods) and a microwave
$\pi$-pulse applied on the $\ket{0}\leftrightarrow\ket{1}$
resonance. During the $\pi$-pulse of duration $\tau$, we apply
detection light with a variable average photon number $n$. The
incident light measures the atomic state and thereby prevents the Rabi
oscillation, as shown in Fig.~\ref{fig:Zeno}. We model this system as
a coherently driven qubit undergoing on average $\tilde{n}$ projective
measurements that are randomly distributed over $\tau$, and solve the
corresponding Bloch equations (BEs). We include technical
imperfections (preparation, detection and pulse errors) which limit
the maximal (minimal) transfer probability to 0.95 (0.02).
From the measured transfer efficiency (Fig.~\ref{fig:Zeno}), we infer
$\tilde{n}$ for each mean number of incident photons $n$ and, as
expected, observe a linear relationship {$\tilde{n}=(0.37\pm0.02)n$} (inset of
Fig.~\ref{fig:Zeno}). To see how the {maximum knowledge $\Imax $}
is related to $\tilde{n}$, we consider the evolution of
the qubit in the absence of the microwave pulse. The BEs predict a
reduction of coherence by $\exp(-\tilde{n})$, while the equivalent
description introduced before (partial trace over the ``meter'')
predicts $\langle\Psi_1|\Psi_0\rangle$. Thus,
$\langle\Psi_1|\Psi_0\rangle=\exp(-\tilde{n})$, and
%$\Imax\approx 2\tilde{n}$.
$\Imax = f(2\tilde{n})$.  The Zeno measurement therefore yields
$\Imax = \fln((0.74\pm0.04)n)$, in reasonable agreement with the
value $\Imax = \fln(0.62n)$ expected from photonic mode overlap. This
  shows that every single photon incident on the cavity leads to a
  significant state collapse, reducing the atomic coherence by a
  factor of 0.7.

\begin{figure}[htb]
\includegraphics[width=\columnwidth]{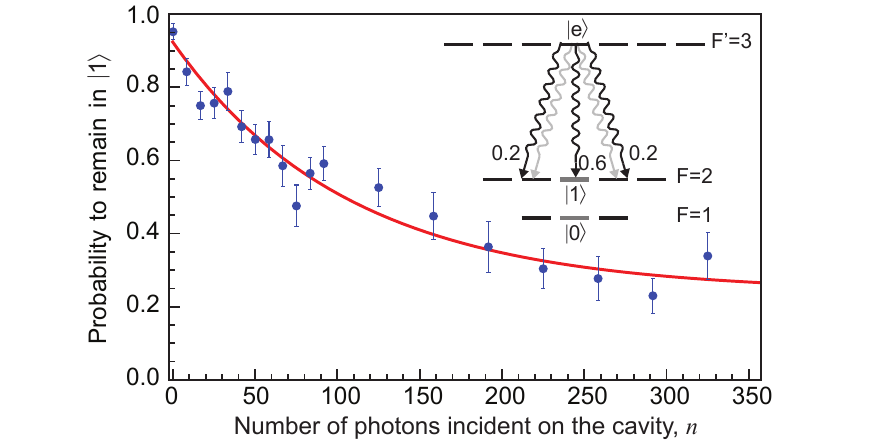}
\caption{\label{fig:QND} \small\textbf{Spontaneous emission during
    detection.} The datapoints (one s.d. error bars) show the measured
  probability that the atom remains in $\ket{1}$ during detection
  versus number of incident photons $n$. The solid line is a fit of an
  exponential decay to a steady-state population of $0.27\pm0.05$ with
  initial rate $\nu=1/(142\pm25)$. The inset shows the two decay
  processes depleting $\ket{1}$. Detection light excites the state
  $\ket{e}$, which can decay into free space (black) with rate
  $\Gamma$ or into the second cavity mode (light grey) with rate
  $\Gamma_P$ (see methods). Correcting for decay back to $\ket{1}$ we
  obtain the number of scattered photons $m=n/(118\pm20)$.}
\end{figure}

In order to relate {the maximum knowledge $\Imax$} and the
accessible {knowledge $\Iacc$} to the number of scattered photons $m$,
we measure $m$ as a function of $n$.  Rather than attempting
direct detection of the spontaneous photons (which would be
inefficient and difficult to calibrate), we take advantage of the
fact that each spontaneous scattering event of the
$\ket{1}=\ket{F=2,m=0}$ state has a known probability to depump to other
states $\ket{F=2,m\neq 0}$. The scattering rate of the
off-resonant state $\ket{0}$ is three orders of magnitude smaller
and can be neglected. We prepare the atom in $\ket{1}$ and turn on
detection light for a variable time. Afterwards, we apply a microwave
$\pi$-pulse on the $\ket{1}\leftrightarrow\ket{0}$ transition, and finally
determine whether the atom has been transferred to $\ket{0}$. The
microwave pulse has no effect on initial states $\ket{F=2,m\neq 0}$,
so that its transfer probability to $\ket{0}$ is equal to the
population remaining in $\ket{1}$ after the detection light pulse. {We
find that this survival probability decays exponentially with the
number of incident photons with an initial rate
$\nu=1/(142\pm25)$ (Fig.~\ref{fig:QND}). Because of the
probability to decay back to $\ket{1}$, the actual spontaneous
emission rate is larger than this depumping rate.} Correcting for this effect (see methods), we obtain
$m/n=1/(118\pm20)$.  This is compatible with the theoretical
prediction {$m/n=1/83$} for our particular atom-cavity system, where the
second cavity mode increases the spontaneous emission rate (see
methods). A still better value, $m/n=\sqrt{T_0}/C$, can be
expected for a single-mode cavity.

\begin{figure}[htb]
  \includegraphics[width=\columnwidth]{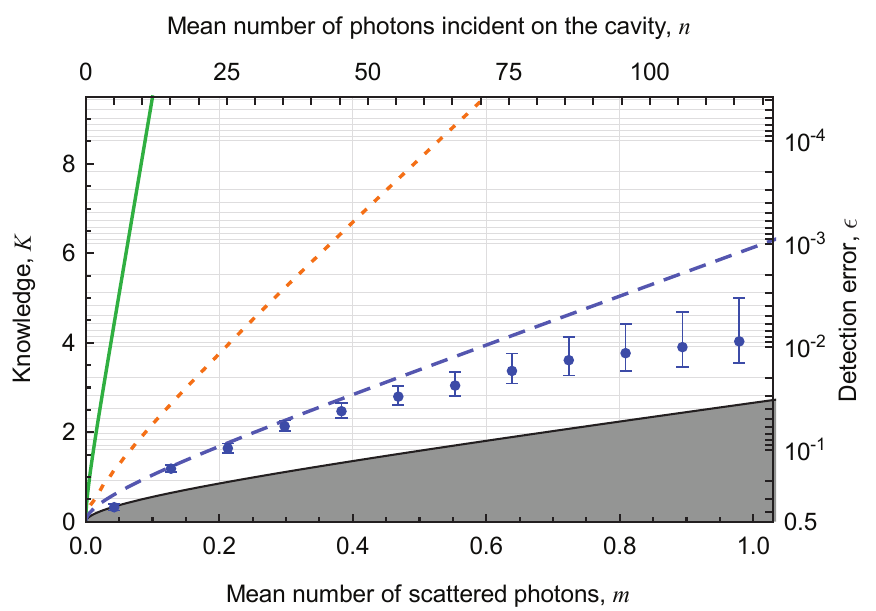}
  \caption{\label{fig:final} \small\textbf{Detection error and knowledge
      versus number of scattered photons.} The grey area is the range
    accessible to free-space detection schemes. This limit is overcome
    using a cavity. Green solid line: maximum knowledge $\Imax$
    extracted by the cavity measurement, deduced from the data in
    figure \ref{fig:Zeno}. Orange dotted line: Accessible knowledge
    using our cavity with perfect photon counters to detect reflected
    and transmitted photons. Blue dashed line: accessible information
    $\Iacc$ with the detection efficiency in our experiment. Blue
    circles ($\bullet$): knowledge actually obtained from the
    experiment with one s.d. error bars. This knowledge is above the
    free-space limit in spite of experimental imperfections. }
\end{figure}

We can now express $\Imax$ and $\Iacc$ in terms of scattered photons
(figure \ref{fig:final}). In the regime $m\ll 1$, where the detection
efficiency is not limited by depumping, we find that our experiment
extracts {$\Imax = \fln((87\pm17) m)$, of which $\Iacc= \fln((5.4\pm0.9) m)$} is
actually accessed.  In spite of experimental imperfections, this {knowledge gain}
is a factor of $2.7$ higher than possible in an ideal fluorescence
measurement and two orders of magnitude larger than in
state-of-the-art experiments \cite{Gerber09}.  Since 118 photons on
average can be sent onto the cavity before one scattering event occurs
and each performs a strong measurement, a large amount of information on the atomic state
can be obtained with negligible scattering. In this sense, one can say
that the photons measure the atom without entering the cavity.

Note that our experiment is still limited by cavity imperfections that
should be straightforward to improve. The closely spaced second cavity
mode is a result of birefringence; experience with macroscopic
cavities suggests that it can be either moved further away or made
degenerate in future cavities. Cavity losses can be further reduced by
at least a factor of four  \cite{Hunger10b} by using state-of-the art
mirror coatings in an otherwise identical fiber
cavity. Assuming these conditions and detector efficiencies of
$70\%$, the accessible {knowledge} would be $\Iacc \approx \fln(110
m)$.

For our cavity parameters, heating mechanisms other than scattering
are expected to be negligible \cite{Domokos03}. Furthermore, the
remaining scattering only leads to a small probability for the atom to
change its vibrational state during detection, because of the
Lamb-Dicke effect in our strong dipole trap. We estimate this
probability to be two orders of magnitude smaller than in
state-of-the-art fluorescence measurements for the same {knowledge}
gain (in our case we can access $\Iacc\approx70$ before the creation
of a phonon).  This allows detection of atomic qubits while staying in
the ground state of the trap, thereby removing the necessity of
recooling after read-out and drastically improving the cycling time of
atom-based quantum computing schemes. Furthermore, the cavity readout
scheme disposes of the requirement for closed transitions in state
readout, opening perspectives for detection of single cold
molecules \cite{Jones06}.

\bigskip

\textbf{Acknowledgements} This work was funded in part by the AQUTE
Integrated Project of the EU (grant no.~247687), by the Institut
Francilien pour la Recherche sur les Atomes Froids (IFRAF), and by the
EURYI grant ``Integrated Quantum Devices''.

\textbf{Autor contributions} J.V., R.G. and G.D. performed the
experiment. All authors contributed to data analysis and
interpretation, as well as to the manuscript.

\textbf{Author information} The authors declare that they have no
competing financial interests. Correspondence and requests for
materials should be addressed to J.~R. (email: jakob.reichel@ens.fr).

\newpage

\setlength{\parindent}{0pt}
\setlength{\parskip}{6pt}

\section*{Methods}
\textbf{Preparation of the qubit states}
We initially extract a single atom in the F=2 hyperfine ground state and unknown Zeeman state from a BEC \cite{Gehr10}. We then apply a microwave $\pi$-pulse on the qubit transition $\ket{0}\leftrightarrow\ket{1}$ followed by a short detection light pulse. If and only if the atom initially is in $\ket{1}$, the $\pi$-pulse transfers it to $\ket{0}$, leading to high cavity transmission. If it is in a different Zeeman sublevel, it remains in $F=2$, leading to low transmission.
In this case, spontaneous scattering due to the read out pulse leads to a redistribution in the $F=2$ multiplet. We repeat the procedure until we detect high transmission, signaling an atom in $\ket{0}$.
If we want to prepare the atom in $\ket{1}$ we apply an additional $\pi$-pulse.

\textbf{Extracted and accessible knowledge from an imperfect cavity}
Assuming a coherent state with amplitude $\sqrt{n}$ incident onto the
cavity, a coherent field builds up in the cavity populating the main
(resonant) and the orthogonally polarised, detuned TEM00 mode. Their
amplitudes depend on the qubit state and each decays via three
channels: transmission, reflection and {losses} at the
mirrors. Thus the outgoing field can be approximated by the tensor
product of six coherent fields with amplitudes {$\alpha_i \,
  \sqrt{n}$}. The subscript $i\in\{0,1\}$ denotes the qubit state,
while $\alpha\in\{t_m,r_m,l_m,t_d,r_d,l_d\}$ identifies the outgoing
mode, $m$ and $d$ respectively designating the main and detuned
mode. The overlap between the two possible light states $\ket{\Psi_0}$
and $\ket{\Psi_1}$ is then $\exp(-\zeta n)$ where $\zeta = \sum_\alpha
|\alpha_0 - \alpha_1|^2$, leading to a maximum knowledge $\fln(\zeta
n)$. In our case, the atom in state $\ket{1}$ is resonant to cavity
and light field while the state $\ket{0}$ is far detuned. Under these
circumstances the phase shift between the states $\ket{\Psi_1}$ and
$\ket{\Psi_0}$ can be neglected and the the amplitude coupling factors
$\{\alpha_i\}$ can be considered real. The power coupling factors in
percent are $\{\alpha_0^2\}=\{12.7,41.4,45.9,0,0,0\}$ for an atom in
$\ket{0}$ and $\{\alpha_1^2\}=\{0.1, 99, 0.4, 0.1, 0.1, 0.4\}$ for an
atom in $\ket{1}$. From these values, we deduce $\zeta = 0.62$.

Using counters to detect the transmitted and reflected intensities,
the detection error is minimised by using a maximum likelihood
estimator (thresholding). The minimal error is $\epsilon_D = (1 - ||
P_0- P_1 ||_1 )/2$ where $P_0$ and $P_1$ are the probability
distributions of detected counts if the qubit is in $\ket{0}$ or
$\ket{1}$, and the distance $|| P_0- P_1 ||_1$ is defined as
$\sum_\mathbf{x} | P_0(\mathbf{x})-P_1(\mathbf{x})|/2$, where the sum
goes over all possible detection events. In our case, the two
distributions are the products of two Poisson distributions and we
have numerically observed that their distance is well approximated by
$\sqrt{1 - Q}$, where $Q$ is the Chernoff coefficient
 $Q = \min\limits_{0\leq s\leq1}\left[
\sum_\mathbf{x} P_0^s(\mathbf{x}) P_1^{1-s}(\mathbf{x})\right]$,
which can be calculated to be $Q=\exp(-\xi n)$, where\\
 $\xi = -\min\limits_{0\leq s\leq1} \left[T_0^s T_1^{1-s} + R_0^s R_1^{1-s} -
s (T_0+R_0) - (1-s)(T_1+R_1)\right]$. The accessible knowledge is
then given by the same expression $\fln(\xi n)$ as the {maximum
knowledge} derived from the Helstrom error bound (but of
  course with a different $\xi$), allowing direct comparisons between
the two. In our regime of parameters, the minimum is reached for
$s \approx 0.5$ leading to $\xi = (T_0+T_1+R_0+R_1)/2 - \sqrt{T_0
  \, T_1} - \sqrt{R_0 \, R_1} = 0.11$. Taking into account the finite
detector efficiencies, the same calculation leads to $\xi =
4.6\times10^{-2}$.

\textbf{Determination of the spontaneous scattering rate}
The detection light populates the excited state $\ket{F'=3, m_F=0}$
which can decay via two channels: spontaneous emission into free space
with rate $\Gamma$, and into the second orthogonally polarised TEM00
cavity mode which is detuned by $540$ MHz \cite{Gehr10} with a
Purcell-enhanced rate $\Gamma_P$ (see fig \ref{fig:QND}). Decay into
the original (pumped) cavity mode is not considered since it
constitutes a coherent process which does not change the atomic
state. The decay into the second cavity mode always leads to a change
of the atomic Zeeman state, while for the decay into free space the
atom has a probability of 3/5 (given by the transition strengths) to
end up in the original Zeeman ground state. Therefore, the total
spontaneous emission rate $\Gamma+\Gamma_P$ is higher than the
measured decay constant of state $\ket{1}$ (see
Fig. \ref{fig:QND}). To correct for this small effect we have to know
the relative probability of the two decay channels. The effect of the
second cavity mode is too strong to be treated as a perturbation, therefore
we numerically solve the complete master equation (including all
ground and excited state Zeeman levels as well as the two cavity
modes). From this solution we obtain the ratio of the two decay
channels of $\Gamma_P/\Gamma=2.6$ which only weakly depends on
experimental parameters. Using this value we obtain for the
spontaneous scattering rate {$n/m=\nu^{-1}\times
  (\Gamma_P+2\Gamma/5)/(\Gamma_P+\Gamma)$}. This ratio is 3.6 times
smaller than the value $C/\sqrt{T_0}$ for a single-mode cavity.

\end{document}